\documentclass[aps,twocolumn, floats,preprintnumbers,showpacs,nofootinbib,prd]{revtex4}
\usepackage{graphicx}
\usepackage{url}
\usepackage{amsmath}
\usepackage{amsfonts}
\usepackage{amssymb}
\def\beq{\begin{equation}}
\def\eeq{\end{equation}}
\def\beqa{\begin{eqnarray}}
\def\eeqa{\end{eqnarray}}

\def\ltap{\ \raise.3ex\hbox{$<$\kern-.75em\lower1ex\hbox{$\sim$}}\ }
\def\gtap{\ \raise.3ex\hbox{$>$\kern-.75em\lower1ex\hbox{$\sim$}}\ }

\begin{document}
\preprint{SLAC-PUB-15221} 
\preprint{UCI-HEP-TR-2012-11}

\title{Searches with Mono-Leptons
}

\author{Yang Bai$^{a,b}$ and Tim M.P. Tait$^{c}$
\\
\vspace{2mm}
${}^{a}$Department of Physics, University of Wisconsin, Madison, WI 53706, USA \\
${}^{b}$SLAC National Accelerator Laboratory, 2575 Sand Hill Road, Menlo Park, CA 94025, USA \\
${}^{c}$Department of Physics and Astronomy,
University of California, Irvine, CA 92697, USA
}


\begin{abstract}
We explore the implications of the mono-lepton plus missing transverse energy signature at the LHC, and point out its significance on understanding how 
dark matter interacts with quarks, where the signature arises from dark matter pair production together with a leptonically decaying $W$ boson  radiated from the initial state quarks. We derive limits using the existing $W^\prime$ searches at the LHC, and find an interesting interference between the contributions from dark matter couplings to up-type and down-type quarks. Mono-leptons can actually furnish the strongest current bound on dark matter interactions for axial vector (spin-dependent) interactions and iso-spin violating couplings.  Should a signal of dark matter production be observed, this process can also help disentangle the dark matter couplings to up- and down-type quarks.
\end{abstract}
\maketitle

\noindent
{\it{\textbf{Introduction.}}}
Observational evidence points to the existence of some kind of
cold nonbaryonic dark matter as the dominant component of matter
in the Universe \cite{Bertone:2004pz},  and yet, from the point of view of a
fundamental description, essentially nothing is known about the nature
of dark matter.  Among the many possibilities, weakly interacting massive
particles (WIMPs) are the most cherished vision for dark matter, because
their abundance in the Universe may be simply understood as a consequence
of the thermal history.  But even in the space of WIMP theories, there is a large
set of possible interactions with the ordinary particles of the Standard Model
(SM), leading to a rich program of searches for WIMPs indirectly through their
annihilation, directly scattering with heavy nuclei, and through their production
at high energy accelerators.

If the particles mediating the WIMP interactions with the SM are heavy compared to
the momentum transfer of interest, the ultraviolet details become unimportant, and
low energy physics is described by an effective field theory (EFT)
containing the SM,
the WIMP, and contact interactions coupling the two 
sectors \cite{Beltran:2008xg,Beltran:2010ww,Cao:2009uw,Goodman:2010yf,Bai:2010hh}.
The effective theory has proven a useful language to describe some kinds of
WIMP theories, and assess the interplay of direct searches with those at
colliders \cite{Beltran:2010ww,Cao:2009uw,Goodman:2010yf,Bai:2010hh,Goodman:2010ku,Fox:2011fx,Rajaraman:2011wf}
and indirect detection \cite{Goodman:2010qn,Cheung:2010ua}.  A picture emerges in which the various classes of searches exhibit a high degree of complementarity in terms of their coverage of different theories of WIMPs.

Currently the most sensitive accelerator searches look for mono-jets
and mono-photons which recoil against a pair of invisible WIMPs
\cite{Aaltonen:2012jb,Chatrchyan:2012pa,Chatrchyan:2012tea,atlas-dm}.  In general, the collider searches tend to provide better coverage for spin-dependent interactions and for low mass ($\lesssim 10$~GeV) WIMPs.  In this article, we explore the signature where a ``mono-$W$" boson is produced in association with the WIMPs. When the $W$ decays leptonically, this results in a charged lepton and a neutrino, leading to events characterized by
a single charged lepton and missing transverse momentum 
(see Fig.~\ref{fig:feynman}).  As we shall see below, the existing $W^\prime$
searches already place a bound on mono-$W$ production which for some
choices of couplings are currently the most stringent, better than existing
mono-jet bounds. Even in cases where the mono-leptons do not provide the most stringent constraints, they are an interesting mechanism to disentangle WIMP couplings to up-type versus down-type quarks.

\begin{figure}[!]
\begin{center}
\hspace*{-0.75cm}
\includegraphics[width=0.48\textwidth]{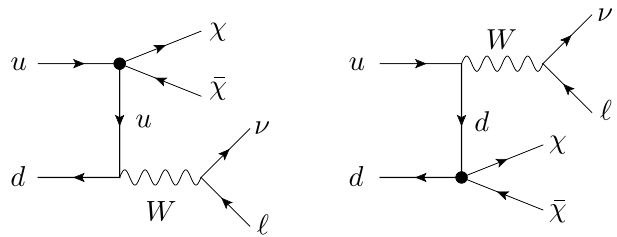}
\caption{Representative Feynman diagrams for $W \chi \bar{\chi}$ production. }
\label{fig:feynman}
\end{center}
\end{figure}

\vspace*{0.3cm}
\noindent
{\it{\textbf{Effective Field Theory.}}}
We consider a theory of a Dirac (electroweak singlet) WIMP particle $\chi$
which interacts with up ($u$) and/or down ($d$) quarks through either a vector
or axial-vector interaction.  The vector case is represented by the contact 
interaction,
\begin{eqnarray}
\frac{1}{\Lambda^2} ~ \overline{\chi} \gamma_\mu \chi ~
\left(
\overline{u} \gamma^\mu u + \xi ~ \overline{d} \gamma^\mu d
\right) ~,
\label{eq:vec}
\end{eqnarray}
where $\Lambda$ characterizes the over-all strength of the interaction,
$\xi$ parameterizes the relative strength of the coupling to down quarks
relative to up-quarks, and for simplicity we restrict our discussion to quarks of
the first generation.  This interaction leads to spin-independent scattering with
nuclei.  We also consider a spin-dependent case with an axial vector structure,
\begin{eqnarray}
\frac{1}{\Lambda^2} ~ \overline{\chi} \gamma_\mu \gamma_5 \chi ~
\left(
\overline{u} \gamma^\mu \gamma_5 u + \xi ~ \overline{d} \gamma^\mu \gamma_5 d
\right)~.
\label{eq:avec}
\end{eqnarray}
where the parameters $\Lambda$ and $\xi$ should be understood as analogous
to but distinct from the vector interaction case.

Because $u$ and $d$ quarks are part of the same $SU(2)_L$ multiplet, one
would naively expect the couplings to show correlations between the two flavors
as well as between the vector and axial-vector structures.  
However, one can upset these naive
expectations by invoking higher dimensional operators containing Higgs insertions
to craft any combination of vector and axial vector interactions, as well as any value
of $\xi$ one likes.  We thus consider some simple representative choices which
isolate the vector or axial-vector structure, as well as different values of $\xi$, below.

\vspace*{0.3cm}
\noindent
{\it{\textbf{Simulation and Results.}}}
We simulate production of $p p \rightarrow \chi \bar{\chi} W$, followed by
$W \rightarrow \ell \nu$ with $\ell = e$ or $\mu$ at the 7 TeV LHC, using
Madgraph 5, with parton showering and hadronization by Pythia, and
process the events with the PGS CMS detector simulation \cite{Alwall:2011uj}.
We find that for $\xi = +1 (-1)$, the production rate shows rather strong
destructive (constructive) interference
between the two diagrams of Figure~\ref{fig:feynman}, the degree of which 
depends on the specific kinematics considered.

We derive limits based on the CMS $W^\prime$ search for a single energetic
lepton and missing transverse momentum,
based on 5 fb$^{-1}$ of data collected at $\sqrt{s} = 7$~TeV 
\cite{Chatrchyan:2012qk}.  Following the CMS analysis,
events are selected containing
an electron (muon) with $p_T \geq 45~(85)$~GeV isolated from hadronic activity.
The primary cut is in terms of the transverse mass,
\begin{eqnarray}
M_T & \equiv &
\sqrt{2 p_T^\ell p_T^{\nu} ~ \left( 1 - \cos \Delta \phi_{\ell \nu} \right)}
\end{eqnarray}
where $p_T^\nu=E^{\rm miss}_T$ is the missing transverse momentum, and
$\Delta \phi_{\ell \nu}$ is the azimuthal opening angle between the charged lepton transverse momentum direction and $\vec{p}_T^{\,\nu}$.  Events are further required to satisfy $0.4 < p_T^\ell / p_T^\nu < 1.5$ and $\Delta \phi_{\ell \nu} > 0.8 \pi$. Based on no observed excess for any value of the cut on $M_T$,
CMS provides limits on the cross section as a function of the $M_T$ cut.
We find that the analysis requiring $M_T \geq 600$~GeV provides the
most stringent bound over most of the dark matter parameter space, and we 
translate the bound on the cross section into bounds on $\Lambda$ as a function
of $m_\chi$ for $\xi = 1, -1$, and $0$. We also include the latest results from the CMS $W^\prime$ search based on 20 fb$^{-1}$ of data collected at $\sqrt{s} = 8$~TeV~\cite{CMS-wprime-8TeV}. The $M_T \geq 1000$~GeV cut in Ref.~\cite{CMS-wprime-8TeV} will be used for the 8 TeV results. 

\begin{figure}[ht!]
\begin{center}
\hspace*{-0.75cm}
\includegraphics[width=0.48\textwidth]{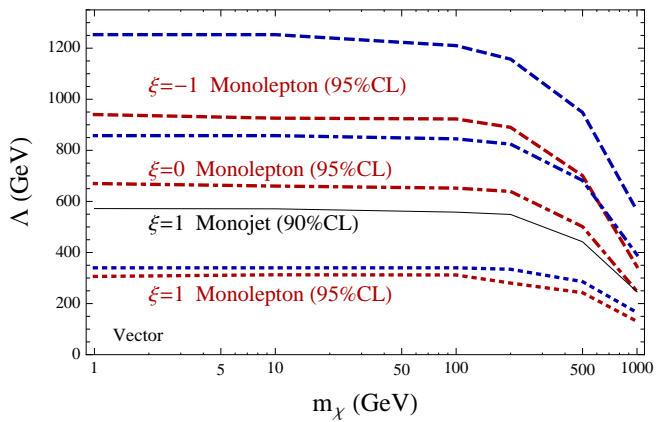}
\caption{The red lines are lower limits on the interaction strength $\Lambda$
(derived from the CMS $W^\prime$ search results at 7 TeV with a 5.0 fb$^{-1}$)
as a function of  the dark matter mass. The dotted, dot-dashed and 
dashed lines are $95\%$ CL limits from the mono-lepton final state for three 
different relations between the up-type and down-type operators: $\xi=1, 0, -1$, 
respectively. The blue lines are constraints from the CMS $W^\prime$ search results at 8 TeV with a 20 fb$^{-1}$. The solid line is the $90\%$ CL CMS limit from mono-jet searches 
with the same luminosity for $\xi=1$. The limits from the mono-jet search 
for $\xi=-1$ are identical to the limits for $\xi=1$, while the limits for $\xi=0$ is only 
slightly weaker.}
\label{fig:cutoff-vector}
\end{center}
\end{figure}

\begin{figure}[ht!]
\begin{center}
\hspace*{-0.75cm}
\includegraphics[width=0.48\textwidth]{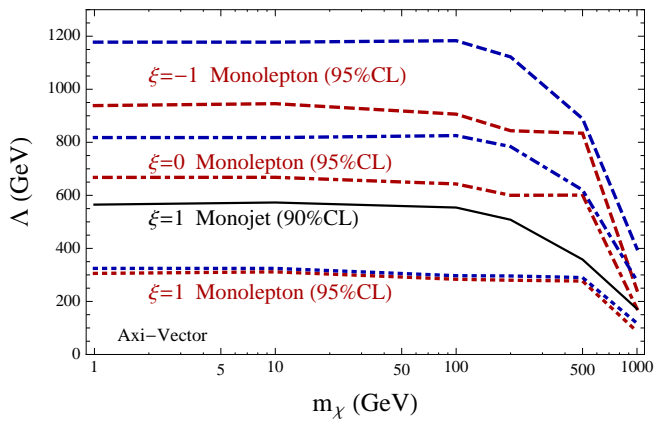}
\caption{The same as Fig.~\ref{fig:cutoff-vector} but for the axial-vector 
operators of Eq.~(\protect{\ref{eq:avec}}). }
\label{fig:cutoff}
\end{center}
\end{figure}

In Figures~\ref{fig:cutoff-vector} and \ref{fig:cutoff}, we present the mapping of the
$95\%$ confidence level (CL) CMS limits on anomalous production of mono-leptons into bounds on $\Lambda$
for the vector and axial-vector interactions, respectively,
as a function of the dark matter mass.  We have chosen three ratios of the
coupling to down-quarks compared to up-quarks, $\xi = 1, 0, -1$ to illustrate
the importance of interference between the two Feynman graphs.  Also shown
for comparison are the $90\%$ CL limits from CMS based on their
dedicated mono-jet search \cite{Chatrchyan:2012pa}
(very similar results have also recently been reported by the ATLAS
collaboration \cite{atlas-dm} and exceed the CDF limits \cite{Aaltonen:2012jb}).  Lacking interference effects, the mono-jet searches provide the same limits for the $\xi=1$ and $\xi=-1$ cases and a slightly weaker limit for the $\xi=0$ case. For $\xi = 1$, the mono-jet search yields
a better limit by roughly a factor of two on $\Lambda$.  For $\xi = 0$, the
mono-lepton search is slightly more restrictive, and for $\xi = -1$ it is
substantially better.  

Comparing the limits from the 7 TeV and 8 TeV data, one can see that a higher center-of-mass energy and a higher luminosity provide better constraints. The future data from the 14 TeV LHC is envisioned to provide potentially better limits by increasing the cut on $M_T$. Since the current best limit from 8 TeV is around 1.2 TeV, which is far below the 14 TeV center of mass energy, special care should be taken to have a consistent effective field theory description either by including the mass-on-shell heavier particles in the productions or estimating the uncertainties of just using the effective operators.  

\begin{figure}[!]
\begin{center}
\hspace*{-0.75cm}
\includegraphics[width=0.48\textwidth]{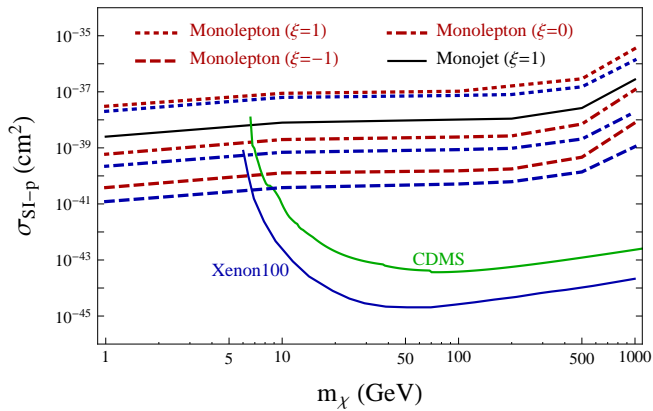}
\caption{Mono-lepton bounds and bounds from direct detection projected
into the plane of the WIMP mass and the spin-independent cross section
with protons. The red and blue lines are from CMS 5 fb$^{-1}$ data at 7 TeV and 20 fb$^{-1}$ data at 8 TeV, respectively.}
\label{fig:SI-p}
\end{center}
\end{figure}

Mapping the bounds into the parameter space of direct detection, in 
Figure~\ref{fig:SI-p} we show the collider limits in the plane of the spin-independent cross section for scattering off protons.  For reference, we have also plotted the recent bounds from Xenon 100~\cite{Xenon200:2012nq} and CDMS~\cite{Ahmed:2009zw} which assume $\xi=1$. 
For  $\xi=0, -1$, the Xenon and CDMS limits are rescaled from the $\xi=1$ values
by the order one fractional proton content of the isotopes of Xenon and Germanium,
respectively.  
As is typical, collider bounds represent the best existing limits for very low WIMP masses ($m_\chi \lesssim 7$~GeV), where WIMPs in the galactic halo typically have too little momentum to register in conventional direct detection experiments. For $\xi = 0,-1$, the mono-lepton bounds are currently the world's best for such low mass WIMPs.

\begin{figure}[t!]
\begin{center}
\hspace*{-0.75cm}
\includegraphics[width=0.48\textwidth]{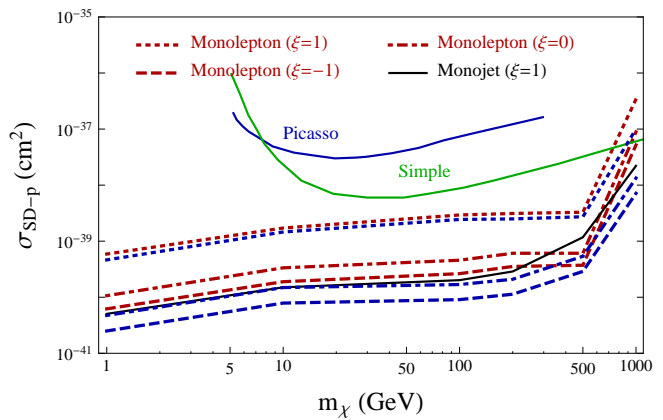}
\caption{Mono-lepton bounds and bounds from direct detection projected
into the plane of the WIMP mass and the spin-dependent cross section
with protons. The red and blue lines are from CMS 5 fb$^{-1}$ data at 7 TeV and 20 fb$^{-1}$ data at 8 TeV, respectively.}
\label{fig:SD-p}
\end{center}
\end{figure}

\begin{figure}[t!]
\begin{center}
\hspace*{-0.75cm}
\includegraphics[width=0.48\textwidth]{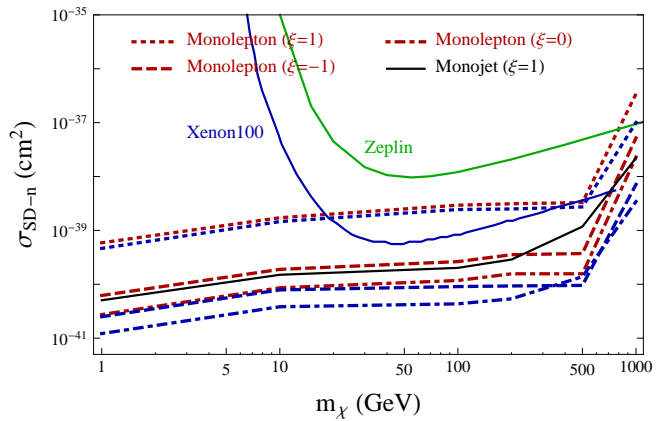}
\caption{Mono-lepton bounds and bounds from direct detection projected
into the plane of the WIMP mass and the spin-dependent cross section
with neutrons. The red and blue lines are from CMS 5 fb$^{-1}$ data at 7 TeV and 20 fb$^{-1}$ data at 8 TeV, respectively.}
\label{fig:SD-n}
\end{center}
\end{figure}

In Figures~\ref{fig:SD-p} and \ref{fig:SD-n}, we show the mapping of the axial
vector interaction into the space of the spin-dependent cross section for scattering
off of protons and neutrons, respectively.  For reference, spin-dependent
bounds from Xenon-100 \cite{Xenon100n}, Zeplin-III \cite{Akimov:2011tj}, 
PICASSO \cite{Archambault:2012pm}, and SIMPLE \cite{Felizardo:2011uw} are
also shown.  In the case of spin-dependent interactions, colliders are typically
more sensitive probes for a wide range of masses, losing sensitivity only for
large ($\sim$~TeV) WIMP masses which are difficult to produce relativistically
at LHC energies.  Again, for $\xi=1$ the bounds from mono-jet searches are
typically providing stronger bounds than mono-leptons, but for $\xi =0$ or $-1$,
the repurposed mono-lepton search provides somewhat stronger constraints.

\vspace*{0.3cm}
\noindent
{\it{\textbf{Discussion and Outlook.}}}
We have examined the signal of mono-$W$s, decaying into mono-leptons,
as a means to study WIMP interactions with quarks at the LHC.  This signal was
previously appreciated as a $W^\prime$ search, but we show that it can also provide, in some cases, the most sensitive probe of theories of dark matter.
To evaluate the effectiveness of this search strategy, we have repurposed an existing CMS search for $W^\prime \rightarrow \ell \nu$, and used it to produce bounds on the  interaction strength of WIMPs with quarks for both vector and axial-vector interactions. 
Compared to the mono-jet searches, the mono-lepton searches are cleaner with smaller experimental systematic errors, and are likely to scale better 
than mono-jet searches with increased
luminosity and/or pile-up.
We find that the rate of WIMP + $W$ production is very sensitive to the relative sign of the WIMP coupling to up or down quarks, and mono-lepton searches can provide the best current limits depending on the relative strength and sign of the up- and down-quark interactions.  Should a positive WIMP signal be discovered, the mono-lepton channel provides a key sensitive foil which helps discriminate up and down couplings, including the relative sign between the two. For a light WIMP, the positive results from mono-jet searches can be used to determine the scale of the cutoff. The positive results from mono-lepton can then be used to determine the ratio of the up- and down-quark couplings. A cross-check can be performed at direct detection experiments with different target nuclei. 

Ultimately, whether or not effective field theories prove fruitful as a description of
dark matter production at colliders will depend on the masses of the particles
mediating the interactions.  For the particular search at hand, this currently
would imply that the masses of such particles should be larger than roughly
the cut on $M_T$, and thus the EFT should provide a reasonably accurate description
even for weakly coupled particles.  Nonetheless, even a break-down of the EFT
provides interesting information.  For example, a positive signal at a direct detection
experiment combined with a null result at colliders already would suggest a light
mediator, and help devise more targeted searches to probe it directly~\cite{Bai:2010hh}.

Mono-leptons are an interesting, clean hadron collider 
signature, and one which may ultimately prove 
effective at searches far beyond looking for $W^\prime$s and can be recast for other new physics searches~\cite{Cranmer:2010hk,recastnode}.  Getting the most out of this signature may involve retuning the analysis slightly to maximize sensitivity, and we have shown that this would be a worthwhile exercise for the LHC experiments, given the strong sensitivity to some theories of dark matter.
And dark matter is only one item on a list of well-motivated models which mono-leptons can bound or reveal. For example, in theories of large extra dimensions \cite{ArkaniHamed:1998rs}, mono-leptons can arise in processes where a $W$ boson is produced together with a KK graviton \cite{Balazs:1999ge} and
they also arise in theories with non-standard neutrino-quark 
interactions \cite{Chivukula:1987wf,Friedland:2011za} and could prove useful to search for SUSY models with compressed spectra~\cite{LeCompte:2011fh}. 
We look forward to seeing their full potential explored.

\vspace{3mm}
{\it{\textbf{Acknowledgements.}}}
We thank Roni Harnik, JoAnne Hewett,
and Daniel Whiteson for useful discussion. SLAC is operated by Stanford University for the US Department of Energy under contract
DE-AC02-76SF00515. TMPT acknowledges the
hospitality of the SLAC theory group, and is supported in part by NSF
grant PHY-0970171. We also thank the Aspen Center for Physics, under NSF Grant No. 1066293, where part of this work was completed.

\end{document}